\documentclass[11pt,a4paper]{article}
\usepackage{mathrsfs}
\usepackage{cite}
\usepackage{subfigure}
\usepackage{graphicx}
\DeclareFontFamily{U}{wncy}{}
\DeclareFontShape{U}{wncy}{m}{n}{<->wncyr10}{}
\DeclareSymbolFont{mcy}{U}{wncy}{m}{n}
\DeclareMathSymbol{\Sh}{\mathord}{mcy}{"58}

\newtheorem{theorem}{Theorem}[section]

\usepackage[cmex10]{amsmath}
\usepackage{amssymb}  
\usepackage{algorithmic}

\begin{document}
\title{Estimating the Signal Reconstruction Error from Threshold-Based Sampling Without Knowing the Original Signal}

\author{Bernhard A. Moser\\
 SCCH, Austria\\
Email: bernhard.moser@scch.at }

\maketitle
\begin{center}
{\bf Abstract}
\end{center}
\noindent
The problem of estimating the accuracy of signal reconstruction from threshold-based sampling, by only taking the sampling output into account, is addressed. The approach is based on re-sampling the reconstructed signal and the application of
a distance measure in the output space which satisfies the condition of quasi-isometry.
The quasi-isometry property allows to estimate 
the reconstruction accuracy from the matching accuracy between the sign sequences 
resulting from sampling and the re-sampling after reconstruction.
This approach is exemplified by means of leaky integrate-and-fire. 
It is shown that this approach can be used for parameter tuning for optimizing the 
reconstruction accuracy.
\\
\\
\noindent
Threshold-based Sampling, Signal Reconstruction, Quasi-Isometry, Discrepancy Norm  


\section{Motivation}
\label{s:motivation}
The quality of signal reconstruction depends basically on three factors: 
a) the theoretical accuracy of the reconstruction algorithm for a specified class of input signals, 
b) the proper choice and adaption of configuration parameters of the reconstruction algorithm, and 
c) numerical problems related from quantization and truncation effects. 

There is a substantial difference between uniform and threshold-based sampling. 
Once the grid of equidistant points in time is fixed uniform sampling becomes a linear mapping from the input space 
of analogue signals to the space of discrete sequences of samples. 

Under reasonable mathematical conditions such as compactness of the time interval and 
bandlimitedness, the resulting sequences of uniform samples from two input signals get as close as required in the max-norm  if the input signals are sufficiently close to each other in the max-norm. 

For threshold-based sampling this continuity property of the sampling operator is not valid any longer~\cite{MoserTSP2014}.
In the end, by threshold-based sampling not only we are losing  
linearity due to the inherent non-linearity of the thresholding operation but also continuity.
It is the lack of continuity of the sampling operator 
which poses a special mathematical challenge when measuring the accuracy of approximations. 

We tackle this challenge by taking up the concept of quasi-isometry, Section~\ref{s:quasi}.
Without referring explicitly to quasi-isometry,  this property was already shown for the threshold-based sampling variants send-on-delta (SOD) and integrate-and-fire (IF) in~\cite{MoserTSP2014}. For an extended analysis of quasi-isometry for threshold-based sampling see~\cite{Moser17}.
In this paper apply this result to leaky integrate-and-fire (LIF) for arbitrary leak parameter $\alpha$, 
Section~\ref{ss:LIF}, in order to show how the proposed approach can be used for optimizing signal reconstruction and estimating its accuracy. 

First of all, we start in Section~\ref{s:intro} with fixing mathematical notation, recalling the concept of quasi-isometry as well as
the concepts of  SOD and LIF sampler along with a sketch of the approximate signal reconstruction method due to~\cite{Feichtinger2012}.

\section{Preliminaries}
\label{s:intro}
\subsection{Mathematics of Distances}
\label{ss:distances}
First of all, let us fix some notation. $1_{I}$ denotes the indicator function of the set $I$, i.e., $1_I(t)=1$ if $t\in I$ and $1_I(t)=0$ else.  
 $\|.\|_{\infty}$ denotes the uniform norm, i.e., $\|f-g\|_{\infty} = \sup_{t \in X} |f(t)-g(t)|$, where $X$ is the domain 
of $f$ and $g$. If $M$ is a discrete set then $|M|$ denotes the number of elements. If $I$ is  an interval, then 
$I$ denotes its length. $\mathcal{I}$ denotes the family of real intervals.

In this section we recall basic notions related to distances such as semi-metric, isometry and quasi-isometry, see
 e.g.,~\cite{EncyclopediaofDistances2009}.

Let $X$ be a set. A semi-metric $d:X\times X \rightarrow [0,\infty)$ is characterized by a) $d(x,x)=0$  for all $x\in X$, b) $d(x,y)=d(y,x)$ for all $x,y \in X$ and c) the triangle inequality
$d(x,z)\leq d(x,y) + d(y,z)$ for all $x,y,z \in X$. 
$d$ is a metric if, in addition to a) the stronger condition a')
$d(x,y)=0$ if and only if $x=y$, is satisfied.
The semi-metric  $\tilde d$ is called {\it equivalent} to $d$, in symbols $d \sim \tilde d$, if and only if there are constants 
$A_1, A_2>0$ such that
\begin{equation}
\label{eq:norm-equivalencecondition}
A_1 d(x,y)  \leq \tilde d(x,y) \leq A_2 \, d(x,y)
\end{equation}
 for all $x$, $y$ of the universe of discourse.

A map $\Phi: X\rightarrow Y$ between a metric space ${(X,d_{X})}$ and another metric space  
$(Y,d_{Y})$ is called {\it isometry} if this mapping is distance preserving, i.e., for any 
$x_1,x_2 \in X$ we have $d_{X}(x_1,x_2) = d_{Y}(\Phi(x_1), \Phi(x_2))$.

The concept of {\it quasi-isometry}  relaxes the notion of  isometry by imposing only a 
coarse Lipschitz continuity and a coarse surjective property of the mapping. 
$\Phi$ is called a {\it quasi-isometry} from 
$(X,d_{X})$ to $(Y,d_{Y})$ if there exist constants 
$A\geq 1$, $B\geq 0$, and $C\geq 0$ such that the following two properties hold:
\\
\noindent
i) For every two elements $x_1, x_2\in X$, the distance between their images is, up to the additive constant $B$, within a factor of $A$ of their original distance. This means,  
$\forall x_1, x_2\in X$
\begin{equation}
\label{eq:quasi1}
{\frac{1}{A}}\,d_{X}(x_1,x_2)-B 
 \leq d_{Y}(\Phi(x_1),\Phi(x_2))  
 \leq A\,d_{X}(x_1,x_2)+B. 
\end{equation}
\\
\noindent
ii) Every element of $Y$ is within the constant distance $C$ of an image point, i.e.,
\begin{equation}
\label{eq:quasi2}
\forall y \in Y:\exists x\in \mathcal{F}:d_{Y}(y,\Phi(x))\leq C.
\end{equation}

Note that for $B=0$ the condition (\ref{eq:quasi1}) reads as Lipschitz continuity condition of the operator $\Phi$.
This means that (\ref{eq:quasi1}) can be interpreted as a relaxed bi-Lipschitz condition.
The two metric spaces $ (X,d_{X})$ and $(Y,d_{Y})$ are called {\it quasi-isometric} if there exists a quasi-isometry $Q$ from $ (X,d_{X})$ to $ (Y,d_{Y})$.

\subsection{Leaky Integrate-and-Fire (LIF)}
\label{ss:LIF}
Given a threshold $\vartheta>0$, a positive constant $\alpha>0$ and an integrable input signal $f$, {\em leaky integrate-and-fire sampling} (LIF) triggers an ``up'' or ``down'' pulse at instant $t = t_{k+1}$ depending on whether the integral 
\begin{equation}
\label{eq:LIF}
F(t_k,t):=\int_{t_{k}}^t f(s) e^{\alpha (s- t)} ds
\end{equation}
 crosses the level $\pm \vartheta$. 
LIF is a well known simplified model of a neuron and is used for simulation purposes in computational neuroscience
~\cite{Gerstner2002}.
The exponential term under the integral can be interpreted as exponential fade of history which 
downgrades the influence of information in the past according to the exponential law.
Note that the larger $\alpha$ the stronger the downgrade and the asymmetry between present and past time.
With $\alpha=0$  the time asymmetry vanishes and all points in time are treated equally.

(\ref{eq:LIF}) is closely related to send-on-delta sampling (SOD), which is the simplest variant of 
threshold-based sampling as it just relies on the comparison of a difference by setting 
\begin{equation}
\label{eq:tildeF}
\widetilde F(t):=F(t, -\infty) 
\end{equation}
and applying the sampling rule 
\begin{equation}
\label{eq:SOD}
\widetilde F(t) - \widetilde F(t_k) = \pm \vartheta.
\end{equation} 

\begin{figure}
  \begin{center}
	      \includegraphics[width=0.75 \columnwidth]{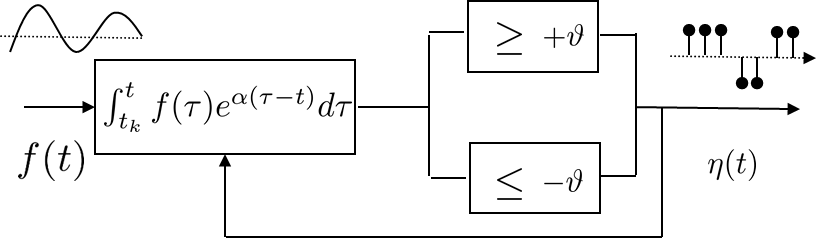} 
  \end{center}
  \caption{Block diagram of LIF according to (\ref{eq:trigger}) with input signal $f$,
	resulting event sequence $\eta(t) := \sum_{k\geq 0} F(t_{k},t_{k+1}) \cdot 1_{\{t_{k+1}\}}(t)$.
	}
  \label{fig:BLOCK}
\end{figure}

\begin{figure}
  \begin{center}
	      \includegraphics[width=0.45 \columnwidth]{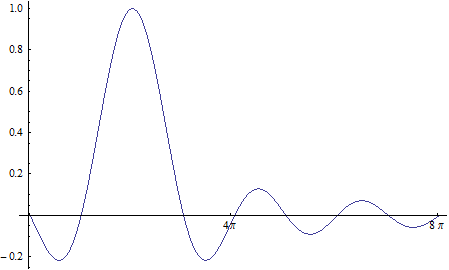}\\
	      \includegraphics[width=0.99 \columnwidth]{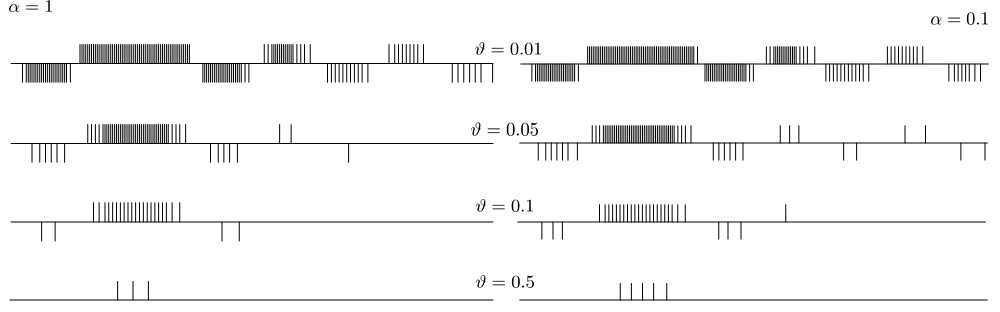} 				
  \end{center}
  \caption{Example of LIF sampled signal with different settings for $\alpha$ and $\vartheta$.
	Bottom left: spike trains resulting from $\alpha = 1$; bottom right: $\alpha = 0.1$.
	The higher $\alpha$ the higher the degree of oblivion, and the higher the threshold $\vartheta$ the sparser
	the encoding by spikes.
	}
  \label{fig:illustrLIF}
\end{figure}

See Fig. \ref{fig:BLOCK} for a block diagram of LIF and Fig. \ref{fig:illustrLIF} for an example with samples resulting from different
settings for $\vartheta$ and $\alpha$.  

The sampling process acts as an operator $\Phi$ that maps a function $f$ to a sequence of pairs $(t_k, v_k)$, 
where $v_k \in \{-\vartheta,\vartheta\}$, $k\geq 1$, represents the up/down-mode of sampling.

Throughout the paper we will assume that the input signals  $f$ are from the Paley-Wiener space $PW_{\Omega}$,
\[
PW_{\Omega}:= \{f \in L^2(\mathbb{R})|\, \mbox{supp}(\hat f) \subseteq [-\Omega, \Omega] \},
\]
where $\hat f(\omega) :=\int_{\mathbb{R}} f(t) e^{-2\pi i \omega t} dt $
is the Fourier transform of $f$.


Let  $\alpha\geq 0$, the threshold $\vartheta>0$ and the signal $f \in PW_{\Omega}$ be given.
The LIF sampler is the recursive process of detecting whether the  evaluation criterion 
\[
\left|\int_{t_{k}}^t f(s) e^{\alpha (s - t)} ds\right| \geq \vartheta
\]
is satisfied, where the detection at instant $t = t_{k+1}$
restarts the process by updating index $k$, i.e.,
\begin{equation}
\label{eq:trigger}
t_{k+1} := \inf\left\{t> t_k \left|\,\,  \left|\int_{t_{k}}^t f(s) e^{\alpha (s - t)} ds\right| \right.\geq \vartheta \right\},
\end{equation}
and where $t_0 = 0$ by definition. 
Let denote by $\tau_{\eta}$ the set $\{t_0, t_1, t_2, \ldots \}$ of time instants $t_k$ with 
$v_k := F(t_{k-1}, t_{k}) \in \{-\vartheta, \vartheta\}$ together with the initial time point $t_0$ and $v_0 = 0$.
In this context, the detection is called {\it event} which  is represented by its time instant $t_k$ and 
the sampling mode $v_k$ at $t_k$, $k\geq 1$. 
 The resulting sequence of events will
 equivalently be represented as sum of its events, i.e., 
\begin{equation}
\label{eq:eta}
\eta(t):= \sum_{k\geq 0}v_k \cdot 1_{\{t_{k}\}}(t).
\end{equation}
\subsection{Approximate Signal Reconstruction from LIF}
\label{ss:Reconstruction}
We recall the approach due to~\cite{Feichtinger2012}. 
Note that 
\[\widetilde F(t) = (f \star \kappa)(t),\]
 $\kappa(t):= e^{-\alpha\, t} 1_{[0,\infty]}$, showing that 
(\ref{eq:tildeF}) can be represented as convolution of $f \in PW_{\Omega}$ with $\kappa \in L^1(\mathbb{R})$, hence
$\widetilde F \in PW_{\Omega}$.
Setting $f(s,t) := f(s) e^{\alpha (s - t)}$, the Leibniz integral rule for variable differentiable integral bound $a(t):= t$, 
\[
\frac{d}{dt}\left(\int_{-\infty}^{a(t)}f(s,t)ds\right) = f(a(t),t) \frac{d}{dt} a(t)  +
\int_{-\infty}^{a(t)} \frac{d}{dt}f(s,t)ds,
\]
yields
\[
\frac{d}{dt} \widetilde F(t) = f(t) - \alpha \widetilde F(t),
\]
hence, in the Fourier domain
\begin{equation}
\label{eq:fDach}
\hat f(\omega) = (2\pi i \omega  + \alpha)\, \hat{\widetilde F}(\omega).
\end{equation}
Choose a Schwartz window function $\psi$ with a) $\hat \psi \equiv 1$ on $[-\Omega, \Omega]$, $\Omega \in (0, \infty)$,
 and b) $\hat \psi$ has compact support.
By replacing  $\hat{\widetilde F}(\omega)$ by its Discrete Time Fourier Transform in 
(\ref{eq:fDach}) and taking 
\[
\hat{\widetilde{F}} = \hat{\widetilde{F}} \star \hat{\psi}
\]
into account we get 
\begin{eqnarray}
\label{eq:oversamplingtrick}
f(t) & = & \sum_{n=-\infty}^{\infty} \widetilde F(n\,T) \, \varphi(t - n \, T) 
\end{eqnarray}
for some $T \in (0, \frac{1}{2\Omega})$,
where $\hat \varphi(\omega):= (2\pi i \omega  + \alpha)\, \hat \psi(\omega)$.
As pointed out by~\cite{Feichtinger2012}, the values 
$\widetilde F(n\,T)$ in (\ref{eq:oversamplingtrick}) 
can be approximated  by values $a_n$ resulting from an algorithm that only uses the LIF samples.
Under the assumptions  
\begin{itemize}
\item $f \in \mbox{PW}_{\Omega}$, and
\item $|\widetilde F(t)| \leq \vartheta$ for all $t \leq 0$
\end{itemize}
the  error bound is given by  $|a_n - \widetilde F(n\, T)| \leq 2\, \vartheta$ for all $n \in \mathbb{Z}$.
The resulting approximation is denoted by $\widetilde f_r$,
\begin{equation}
\label{eq:appr}
\widetilde f_r(t)  =  \sum_{n=-r}^{r} a_n \, \varphi(t - n \, T), 
\end{equation}
where $r$ denotes the truncation index. 
Since for practical and numerical reasons the summation in (\ref{eq:appr}) needs truncation, we will expect
artefacts in the reconstruction result. 
Next we will introduce a method that allows to quantify the reconstruction quality 
in a way that only takes LIF samples  into account.

\section{LIF as Mapping between Metric Spaces}
\label{ss:mapping}
The mapping $f \mapsto \eta$ is denoted by $\Phi$ or $\Phi_{\vartheta}$ in case of emphasizing the threshold $\vartheta$. 
The set of resulting event sequences w.r.t. $\vartheta$ is denoted by 
\begin{equation}
\label{eq:mathcalE}
\mathcal{E} :=  \Phi_{\vartheta}(\mathcal{PW}_{\Omega}).
\end{equation} 
Note that $\mathcal{E}_{\vartheta}$ is subset of the  vector space 
$\big\{
\sum_{k=1}^n a_k\eta_k \,\big|\, 
n \in \mathbb{N},\, a_k \in \mathbb{R},\, \eta_k \in \mathcal{E}_1 \big\}$.
Later on, we will consider metrics in $\mathcal{E}_{\vartheta}$ which are induced by a norm defined in the 
corresponding vector space.

\subsection{Quasi-Isometry for LIF}
\label{s:quasi}
Let  $f \in \mbox{PW}_{\Omega}$, $\Omega>0$, be given.
First of all let us introduce the interval function
\[\mu_f: I \mapsto \mu_f(I) \in \mathbb{R},\]
 for any interval $I \subseteq \mathbb{R}$ by setting
\begin{equation}
\label{eq:LIFmu}
\mu_f((r,s]):= \int_r^s f(\tau) e^{\alpha (\tau - s)} d\tau.
\end{equation}
Further, let us introduce the pseudo-addition $\oplus$ by defining
\begin{equation}
\label{eq:fade}
\mu_f((r,s]) \oplus \mu_f((s,t]) := e^{\alpha(s-t)} \mu_f((r,s]) + \mu_f((s,t]). 
\end{equation} 
It is interesting to observe that $\mu_f$ resembles a pseudo-additive measure in the sense 
that it formally satisfies the generalized additivity condition 
\[
\mu_f(I) \oplus \mu_f(J) = \mu_f(I \cup J)
\]
for disjoint intervals $I \cap J = \emptyset$ with non-empty  intersection of their closures,
$\overline{I} \cap \overline{J} \neq \emptyset$.
Note that the operation $\oplus$ in (\ref{eq:fade}) is associative, i.e.,
\begin{eqnarray}
\label{eq:ass}
 & & \left(\, \mu_f((r,s]) \oplus \mu_f((s,t])\, \right) \oplus \mu_f((t,u]) \nonumber\\
&=& \mu_f((r,s]) \oplus \left(\,\mu_f((s,t]) \oplus \mu_f((t,u])\,\right), 
\end{eqnarray}
for all $r< s < t < u$.

Though introduced by means of the interval function $\mu_f$  
the operation $\oplus$ can also be defined for the discrete case of event sequences  
by exploiting the associativity property (\ref{eq:ass}). 
Given an event sequence $\eta_f(t) = \sum_{k=1}^{\infty} a_k 1_{\{t_k\}}(t)$,
let us define
\begin{eqnarray}
\label{eq:mueta}
\mu_{\eta_f}(I) & := & a_m 1_{\{t_m\}} \oplus \cdots \oplus a_n 1_{\{t_n\}} \nonumber \\
								& := & \sum_{j = m }^n e^{\alpha (t_j - t_n)} a_j, 
\end{eqnarray}
where $I \in \mathcal{I}$ is an interval and
\begin{eqnarray}
\label{eq:tm}
t_m &:=& \min\{t_k \in I|\, \eta_f(t_k) \neq 0 \}, \nonumber \\
t_n &:=& \max\{t_k \in I|\, \eta_f(t_k) \neq 0 \}. \nonumber 
\end{eqnarray}
Now we are able to define the metrics
\begin{eqnarray}
\label{eq:metricF}
d_{\mathcal{F}}(f,g) &:=& D(\mu_f, \mu_g),   \\
\label{eq:metricE}
d_{\mathcal{E}}(\eta_f,\eta_g) &:=& D(\mu_{\eta_f}, \mu_{\eta_g}) . 
\end{eqnarray}
where 
\begin{equation}
\label{eq:D}
D(\mu, \nu):= \sup_{I \in \mathcal{I}}\left| \mu(I) - \nu(I) \right|
\end{equation}
denotes Weyl's discrepancy measure~\cite{Weyl1916}. 
See also~\cite{Chazelle2000} for an overview of discrepancy theory, and~\cite{Moser2012a} for a geometric characterization.

Given $a < b$, suppose that $t_k$ is the last event before $a$ and that there is no event between
$(a,b)$, then 
\[
\left|e^{\alpha (a-b)} \int_{t_k}^a f(\tau) e^{\alpha (\tau - a)} d\tau + \int_{a}^b f(\tau) e^{\alpha (\tau - b)}d\tau
\right| \leq \vartheta
\]
implies
\begin{equation}
\label{eq:ineqvar}
\left| \int_{a}^b f(\tau) e^{\alpha (\tau - b)}d\tau \right| \leq 2 \vartheta.
\end{equation}

\begin{theorem}
\label{th:quasi}
There is a threshold $\vartheta_0>0$ such that for all $0 < \vartheta < \vartheta_0$ 
the mapping 
$\Phi_{\vartheta}: (\mbox{PW}_{\Omega}, d_{\mathcal{F}}) \rightarrow (\mathcal{E}, d_{\mathcal{E}})$
induced by LIF is a quasi-isometry w.r.t. the metrics $d_{\mathcal{F}}$ and $d_{\mathcal{E}}$ defined in (\ref{eq:metricF})
and (\ref{eq:metricE}), respectively.
\end{theorem}

\noindent
Proof.
Since $f,g \in \mbox{PW}_{\Omega}$ we have $d_\mathcal{F}(f,g) < \infty$.
Suppose the non-trivial case $d_\mathcal{F}(f,g)>0$.
Given $\varepsilon >0$, then there is an interval $(a,b]$ such that 
\[
|\mu_f((a,b]) - \mu_g((a,b])| > d_\mathcal{F}(f,g)  - \varepsilon.
\]
Assume without loss of generality $\mu_f((a,b]) > \mu_g((a,b])$.
Define 
\begin{eqnarray}
\label{eq:tr1}
t_b &:= &  \max\{ t_k \leq b \,|\, \max\{|\eta_f(t_k)|, |\eta_g(t_k)|\} > 0 \}, \nonumber\\
\label{eq:tr2}
t^a &:= &  \min\{ t_k \geq a \,|\, \max\{|\eta_f(t_k)|, |\eta_g(t_k)|\} > 0 \}. \nonumber
\end{eqnarray}

Applying (\ref{eq:ineqvar}) on the border intervals $(a, t^a]$ and $(t_b, b]$ yields 
\begin{eqnarray}
\label{eq:trrr}
 &    &  				| \mu_f((a,b]) - \mu_g((a,b])|		\nonumber \\
 & =  &   \mu_f((a,t^a])  \oplus \mu_f((t^a, t_b]) \oplus \mu_f((t_b, b]) -     \nonumber \\
& & 		\mu_g((a,t^a])  \oplus \mu_g((t^a, t_b]) \oplus \mu_g((t_b, b])			   \nonumber \\			
& \leq &    | \mu_f((t^a, t_b]) - \mu_g((t^a, t_b]) | + 8\vartheta_0                   \nonumber\\
& =  &      | \mu_{\eta_f}((t^a, t_b]) - \mu_{\eta_g}((t^a, t_b]) | + 8\vartheta_0     \nonumber\\
& \leq &     d_{\mathcal{E}}(\eta_f, \eta_g) + 8\vartheta_0, \nonumber
\end{eqnarray}
proofing the second inequality of the quasi-isometry condition,
\begin{equation}
\label{eq:D1}
D(\mu_f, \mu_g) \leq D(\mu_{\eta_f}, \mu_{\eta_g}) + 8 \,  \vartheta.
\end{equation}

Now, let us check the first inequality of the quasi-isometry inequality.
Given $\varepsilon>0$. There is an 
interval $[t_m, t_M]$ such that 
$|\mu_{\eta_f}([t_m, t_M]) - \mu_{\eta_g}([t_m, t_M])| > D(\mu_{\eta_f},  \mu_{\eta_g})- \varepsilon$.
Without loss of generality let $\mu_{\eta_f}([t_m, t_M])  > \mu_{\eta_g}([t_m, t_M])$
and $\eta_f(t_m) \neq 0$ and $\eta_g(t_M) \neq 0$.
Consider $t'_m$ the first event of $g$, $f$ after 
$t_m$ and $t'_M$ the last event of $g$, $f$ before $t_M$, i.e., 
$t_m < t'_m < t'_M < t_M$.
Then, again by (\ref{eq:ineqvar}) we obtain the inequalities
\begin{eqnarray}
 & & \mu_{\eta_f}([t_m, t_M]) - \mu_{\eta_g}([t_m, t_M])  \nonumber \\	
& = &  \mu_f((t_m,t'_m])  \oplus \mu_f((t'_m, t'_M])  \oplus \mu_f((t'_M,t_M]) -\nonumber \\	
&   &  \mu_g((t_m,t'_m])   \oplus \mu_g((t'_m,t'_M])  \oplus \mu_g((t'_M,t_M]) \nonumber \\	
& \leq & | \mu_f((t'_m, t'_M]) -\mu_g((t'_m, t'_M]) | + 8\vartheta_0 \nonumber \\
& \leq & D(\mu_f, \mu_g) + 8\vartheta_0 \nonumber,
\end{eqnarray}
proofing
\begin{equation}
\label{eq:D2}
 D(\mu_{\eta_f}, \mu_{\eta_g}) - 8 \,  \vartheta \leq D(\mu_f, \mu_g).
\end{equation}
$\,\,\,\Box$
\section{Measuring the Accuracy of Signal Reconstruction in the Sample Space}
\label{s:accuracy}
Given a LIF sampler with parameter $\alpha>0$ and $\vartheta>0$, 
Theorem~\ref{th:quasi} allows us to estimate the accuracy of signal reconstruction from LIF samples.
Suppose a sequence of events $\eta_a$ given by $\{ a_k 1_{t_k}\}$, $k = 1, \ldots, N$, then a signal reconstruction algorithm under consideration, such as that outlined in~\cite{Feichtinger2012},  reconstructs an approximate signal $\widetilde f$ 
on the time interval $[t_1, t_N]$. By re-sampling the signal $\widetilde f$ we obtain a further sequence of events
$\eta_b$ given by $\{b_k 1_{t_k}\}$, $k = 1, \ldots, M$.  
Since Theorem~\ref{th:quasi} guarantees quasi-isometry in terms of the inequalities~(\ref{eq:D1}) and (\ref{eq:D2}), 
the computation of the discrepancy in the sample space, $D(\eta_a, \eta_b)$, according to (\ref{eq:metricE}) approximates
the discrepancy measure in the signal space, $D(f, \widetilde f)$, according to (\ref{eq:metricF}), where $f$ denotes the 
original input signal. See Fig.~\ref{fig:illustrLIF} for an example. 
This figure shows the reconstruction errors for different truncation parameters measured 
in three different ways: first, the  discrepancy (\ref{eq:metricE}) measured in the sample space; second, 
the discrepancy (\ref{eq:metricF}) in the signal space, and, third, the max-norm between the original and the reconstructed signal.
As expected, the error curves resulting from measuring the discrepancy in the signal and the sample space, respectively,
have similar shapes. In particular, they have approximately the same basin of minimum. 
This means that parameter tuning can also be done in the sample space.

\begin{figure}
  \begin{center}
	      \includegraphics[width=0.75 \columnwidth]{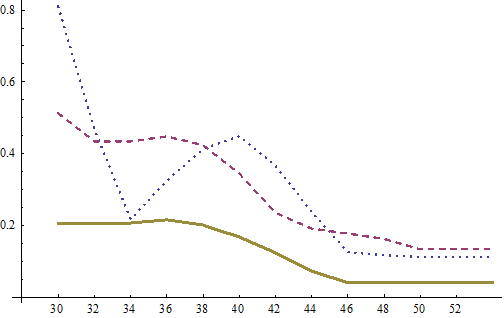} 
  \end{center}
  \caption{Measuring the reconstruction error for the signal of Fig.~\ref{fig:illustrLIF} from LIF samples with the
	parameters $\alpha =0.1$ and $\vartheta = 0.01$; 	the signal is approximately reconstructed due to~\cite{Feichtinger2012} with different truncation indices $r = 30,31,\ldots, 54$; the approximation error is measured in three different ways:
	{\bf solid line}, discrepancy (\ref{eq:metricE}) in the sample space between spike sequences after 
	re-sampling; {\bf dashed line}, discrepancy (\ref{eq:metricF}) in the signal space between input and reconstructed signal;  
		{\bf dotted line}, max-norm in the signal space between input and reconstructed signal.
	}
  \label{fig:errorPlot}
\end{figure}

\section{Conclusion}
It is shown that leaky integrate-and-fire satisfies the condition of quasi-isometry if
the metrics in the input as well as in the output space rely on Weyl's discrepancy measure.
The quasi-isometry relation is utilized for estimating the signal reconstruction error by means of
re-sampling the reconstructed signal. 
An example is presented which demonstrates how numerical errors resulting from truncating summation 
in the reconstruction algorithm can be minimized this way.
In future research we will exploit the outlined quasi-isometry approach as
basis for developing a sound discrete mathematical framework for event-based signal and image processing.


\section*{Acknowledgment}
The author would like to thank the Austrian COMET Program.



%

\end{document}